\documentclass[twocolumn,reprint,superscriptaddress,prb,aps,showpacs,msmath,amsmath,amssymb0]{revtex4-2}
\usepackage{graphicx}
\usepackage{dcolumn}
\usepackage{bm}
\usepackage{subfigure}
\usepackage{natbib} 
\usepackage{multirow}
\usepackage{soul}
\usepackage{float}
\usepackage[normalem]{ulem}
\usepackage[usenames,dvipsnames]{color}
\usepackage{appendix}
\usepackage{mathrsfs}
\usepackage[colorlinks=true,linkcolor=blue,citecolor=blue,urlcolor=blue]{hyperref}

\def\QE{\textsc{Quantum ESPRESSO}\,}

\begin{document}

\title{Comparative study of magnetic exchange parameters and magnon dispersions in NiO and MnO from first principles}

\author{Flaviano José dos Santos}
\affiliation{PSI Center for Scientific Computing, Theory, and Data, Paul Scherrer Institute, 5232 Villigen PSI, Switzerland}
\affiliation{National Centre for Computational Design and Discovery of Novel Materials (MARVEL), 5232 Villigen PSI, Switzerland}
\affiliation{Centro Brasileiro de Pesquisas Físicas (CBPF), Rua Dr. Xavier Sigaud 150, Urca, Rio de Janeiro - RJ, 22290-180, Brazil}

\author{Luca Binci}
\affiliation{Department of Materials Science \& Engineering, University of California Berkeley, Berkeley, CA, 94720, USA}
\affiliation{Materials Sciences Division, Lawrence Berkeley National Laboratory, Berkeley, CA, 94720, USA}
\affiliation{Theory and Simulation of Materials (THEOS), and National Centre for Computational Design and Discovery of Novel Materials (MARVEL), École Polytechnique Fédérale de Lausanne, CH-1015 Lausanne, Switzerland}

\author{Guido Menichetti}
\affiliation{Dipartimento di Fisica dell'Universit\'a di Pisa, Largo Bruno Pontecorvo 3, I-56127 Pisa,~Italy}
\affiliation{Istituto Italiano di Tecnologia, Graphene Labs, Via Morego 30, I-16163 Genova,~Italy}

\author{Ruchika Mahajan}
\affiliation{Department of Chemical Engineering, Stanford University, Stanford, CA 94305, USA}
\affiliation{SUNCAT Center for Interface Science and Catalysis, SLAC National Accelerator Laboratory, Menlo Park, CA 94025, USA}

\author{Nicola Marzari}
\affiliation{PSI Center for Scientific Computing, Theory, and Data, Paul Scherrer Institute, 5232 Villigen PSI, Switzerland}
\affiliation{Theory and Simulation of Materials (THEOS), and National Centre for Computational Design and Discovery of Novel Materials (MARVEL), École Polytechnique Fédérale de Lausanne, CH-1015 Lausanne, Switzerland}

\author{Iurii Timrov} \email[]{ iurii.timrov@psi.ch}
\affiliation{PSI Center for Scientific Computing, Theory, and Data, Paul Scherrer Institute, 5232 Villigen PSI, Switzerland}

\begin{abstract}
Spin-wave excitations are fundamental to understanding the behavior of magnetic materials and hold promise for future information and communication technologies. Yet, modeling these accurately in transition-metal compounds remains challenging, starting from the self-interaction errors affecting localized and partially filled $d$-orbitals in density-functional theory (DFT) with (semi-)local functionals. In this work, we compare three advanced first-principles approaches for computing magnetic exchange parameters and magnon dispersions in NiO and MnO, all based on a common DFT+$U$ ground state with \textit{ab initio} Hubbard $U$ values obtained from density-functional perturbation theory. Two methods extract exchange parameters directly: one via total-energy differences using the four-state mapping ($\Delta E$), and the other via the magnetic force theorem (MFT) using infinitesimal spin rotations. Magnon dispersions are then obtained from a Heisenberg Hamiltonian through linear spin-wave theory (LSWT). The third approach, time-dependent density-functional perturbation theory with $U$ (TDDFPT+$U$), yields magnon dispersions directly from the dynamical spin susceptibility, with exchange parameters fitted \textit{a posteriori}, for comparison, via LSWT. Our results show that TDDFPT+$U$ and the Heisenberg model based on $\Delta E$-derived parameters align well with experimental neutron scattering data, whereas the MFT-based approach shows larger discrepancies, possibly due to some inherent approximations and limitations of the particular implementation used. This study benchmarks the accuracy of state-of-the-art first-principles techniques for spin-wave modeling and contributes to advancing reliable computational tools for the study and design of magnetic materials.
\end{abstract}

\date{\today} 

\maketitle

\section{Introduction}
\label{sec:intro}

Magnetic materials play a pivotal role in a wide range of modern technologies, from data storage and electronic devices to renewable energy solutions.
They hold further promises in spintronics, magnonics, and quantum information~\cite{Spaldin:2010, Yuan:2022}. Noteworthy classes of magnetic materials that have drawn attention include high-temperature superconductors~\cite{Lee:2006, Zhou:2021}, multiferroics~\cite{Spaldin:2010b, Fiebig:2016}, 2D magnetic materials~\cite{Burch:2018, Gibertini:2019}, magnetic topological materials~\cite{Bernevig:2022, Zhang:2023}, and, recently, altermagnets~\cite{Mazin:2022, Smejkal:2022, Smejkal:2022b}. Various experimental techniques, such as inelastic neutron scattering, resonant inelastic X-ray scattering, and spin-polarized electron energy loss spectroscopy are employed to study these materials. Yet, theoretical models are essential for the interpretation of these experiments. Spin model Hamiltonians, such as the Ising, XY, and Heisenberg models~\cite{Friedli:2017} are powerful tools for understanding the microscopic origin of macroscopic properties and describing collective spin excitations and their coupling to phonons~\cite{Delugas:2023} and plasmons~\cite{Sayandip2023}. The Heisenberg model typically incorporates the isotropic two-spin exchange interactions $J_{ij}$ between magnetic sites $i$ and $j$. However, depending on the material, additional terms and interactions -- such as single-ion anisotropy, Dzyaloshinskii-Moriya, and Kitaev interactions -- can be included to account for relativistic effects like spin-orbit coupling~\cite{Li:2021}. To enhance predictive accuracy and better reproduce experimental results, further extensions to the model, such as quantum renormalization factors~\cite{Igarashi:1992, Singh:1989} or higher-order magnetic interactions like four-spin terms from cyclic exchange~\cite{Thouless:1965, Roger:1989} may become necessary.

Developing methods to reliably and accurately determine the magnetic interaction parameters (MIPs) is still an active research field. The experimental determination of MIPs is generally achieved by fitting the measured spin-wave (magnon) dispersions. However, the resulting Hamiltonian is not uniquely defined because it depends on assumptions made about the model, such as the number of couplings and the type of interactions.
In addition, even for the nearest-neighbor isotropic exchange coupling, the values obtained from this method can vary significantly. This variation arises not only from differences in experimental techniques but also from the fitting procedure itself, which depends on the number of $J_{ij}$ parameters considered and the inclusion of additional interactions in the Heisenberg Hamiltonian. Alternatively, MIPs can be computed directly from first principles using several established methods~\cite{Li:2021, Mankovsky:2022}, including the total-energy differences ($\Delta E$) approach~\cite{Sivadas:2015, Fischer:2009, Archer:2011, Whangbo:2003, Xiang:2011, Xiang:2013a, Sabani:2020}, the infinitesimal-rotations method (IRM) based on the magnetic force theorem (MFT)~\cite{Liechtenstein:1987, Katsnelson:2000, Szilva:2023}, and the spin-spiral method that fits spin-spiral energy dispersions using the generalized Bloch theorem~\cite{Halilov:1998, Gebauer:2000, Jacobsson:2013}. Once the Heisenberg Hamiltonian is parametrized using one of these methods, magnon dispersions can be computed via linear spin-wave theory (LSWT)~\cite{Bloch:1930, Slater:1930}. These methods rely on different approximations and are implemented with various technical nuances in different electronic-structure codes. Moreover, they are often applied on top of different descriptions of the material's ground state, leading to significant discrepancies in the computed MIPs, as shown in numerous studies (see, e.g., Refs.~\cite{Fischer:2009, Archer:2011, Gopal:2017}). Therefore, a systematic comparison of these approaches is essential, particularly when applied to the ground state described using the same level of theory. In this work, we focus on the $\Delta E$ and IRM methods, which are described in greater detail in the following, and contrast these with a direct calculation of the spin-spin susceptibility as obtained from Hubbard-corrected time-dependent density-functional perturbation theory (TDDFPT+$U$)~\cite{Binci:2025}.

Two widely used flavors of the $\Delta E$ approach are prevalent in the literature. The first, and historically the earlier method, involves determining MIPs by computing the total energies of different spin configurations in supercells and then solving a system of linear equations to extract MIPs from the total energy differences~\cite{Sivadas:2015, Fischer:2009, Archer:2011}. This approach requires calculating the total energy for $N+1$ spin configurations, which differ by the flipping of spins of some sites with respect to the ground state, with $N$ being the number of distinct MIPs.
This method assumes that the exchange interactions are independent of spin configurations, which might not hold for all materials. Additionally, this approach provides quasi-averaged values of the microscopic interactions~\cite{Sabani:2020}, and the self-consistency convergence can be challenging for some spin configurations. The second method is the four-state mapping analysis (FSMA)~\cite{Whangbo:2003, Xiang:2011, Xiang:2013a, Sabani:2020}. In this approach, a supercell is constructed and four spin states ($\uparrow \uparrow$, $\uparrow \downarrow$, $\downarrow \uparrow$, $\downarrow \downarrow$) are considered, each representing a different spin alignment of the spin pair whose interaction needs to be determined. A key advantage of the FSMA method is its ability to directly compute the derivatives of exchange interactions, thereby enabling the determination of spin-lattice coupling~\cite{Xiang:2011}. However, this method is very sensitive to the supercell size used~\cite{Menichetti:2019, Menichetti:2024}, and thus the convergence of the MIPs with respect to the supercell size must be carefully verified.

The IRM is based on MFT and leads to the Liechtenstein-Katsnelson-Antropov-Gubanov formula~\cite{Liechtenstein:1987, Katsnelson:2000, Szilva:2023}. It involves applying an infinitesimal spin rotation to the magnetic ground state and approximating the total energy change by the change in single-particle energies. IRM assumes that magnetic moments are localized on atoms, and it can only be applied when the system’s Hamiltonian is defined in a basis set of localized orbitals. This method can be implemented using various strategies, such as the Korringa-Kohn-Rostoker (KKR) Green’s function method~\cite{Luders:2001, Papanikolaou:2002}, the tight-binding linear muffin-tin orbital (LMTO) method~\cite{Andersen:1984, Turek:1997}, density-functional theory (DFT)~\cite{Hohenberg:1964, Kohn:1965} with localized basis sets~\cite{Oroszlany:2019, MartinezCarracedo:2023}, or DFT with plane-wave basis sets~\cite{Korotin:2015, He:2021} with subsequent wannierization to construct a tight-binding Hamiltonian using maximally localized Wannier functions (MLWFs)~\cite{Marzari:1997, Marzari:2012}. This latter technique is attractive due to the widespread use of plane-wave basis sets in electronic-structure codes. However, a drawback is the high sensitivity of the resulting interaction parameters to the quality of the atom-centering of the MLWFs. Depending on the system, MLWFs may not be perfectly centered on atoms, leading to significant variations in the computed MIPs~\cite{He:2021, MartinezCarracedo:2023} and loss of symmetry. Furthermore, MIPs can be highly sensitive to various technical details in the IRM implementation~\cite{Korotin:2015, MartinezCarracedo:2023, Solovyev:2021, Solovyev:2024}. Although IRM is widely used for computing MIPs~\cite{Solovyev:1996, Pajda:2001, Wan:2009, Fischer:2009}, its accuracy has been recently questioned~\cite{Solovyev:2021, Solovyev:2024}. Hence, assessing the accuracy of IRM relative to other methods is important.

An alternative approach to determining MIPs is to compute magnon dispersion directly from the dynamical spin susceptibility and fit it using the analytic expression from LSWT. The dynamical spin susceptibility can be calculated using either time-dependent density-functional theory (TDDFT)~\cite{Gross:1985} or many-body perturbation theory (MBPT)~\cite{Hedin:1965}. TDDFT is typically used in the linear-response regime in the frequency domain, assuming a small external magnetic perturbation. Several techniques are available for this purpose, including the Dyson~\cite{Rousseau:2012, Lounis:2011, Buczek:2011b, dosSantosDias:2015, Wysocki:2017, Singh:2019, Skovhus:2021}, Sternheimer~\cite{Savrasov:1998, Cao:2018, Liu:2023}, and Liouville-Lanczos~\cite{Gorni:2018} methods. To model ultrafast spin dynamics or strong perturbations, real-time propagation within TDDFT is employed~\cite{TancogneDejean:2020b}. MBPT offers another route, based on solving the Bethe–Salpeter equation on top of a DFT or $GW$ ground state~\cite{Karlsson:2000, Sasioglu:2010, Muller:2019, Olsen:2021, Kucukalic:2025}. While this approach is accurate, it is significantly more computationally demanding than TDDFT, especially when the latter uses the (computationally simple) adiabatic local spin-density approximation (ALSDA) kernel. However, ALSDA often fails to capture magnon excitations accurately in insulating transition-metal and rare-earth compounds. To address this limitation, more advanced TDDFT exchange-correlation kernels have been developed. Among them, ALSDA+$U$~\cite{Liu:2023, Skovhus:2022, Skovhus:2022b, Binci:2025} has shown great promise. In this method, the Hubbard $U$ parameter -- introduced within the DFT+$U$ framework~\cite{Anisimov:1991, Liechtenstein:1995, Dudarev:1998, Kulik:2006, Kulik:2008} -- plays a critical role, as the computed magnon energies are highly sensitive to its value. A particularly appealing strategy is the TDDFPT+$U$ approach of Ref.~\citenum{Binci:2025}, which uses TDDFPT~\cite{Gorni:2018} in combination with \textit{ab initio} Hubbard $U$ values. These $U$ parameters are computed self-consistently using linear-response theory~\cite{Cococcioni:2005}, reformulated within the framework of density-functional perturbation theory (DFPT)~\cite{Timrov:2018, Timrov:2021, Binci:2023}. This method has demonstrated high accuracy across a wide range of materials~\cite{Timrov:2020c, Mahajan:2021, Mahajan:2022, Timrov:2022c, Timrov:2023, Binci:2023, Gebreyesus:2023, Bonfa:2024, Gelin:2024, Grassano2024, Haddadi:2024, Macke:2024, Chang:2025, Uhrin:2025, Kam:2025, Bastonero:2025}, and holds strong potential for accurately and efficiently describing magnetic excitations from first principles.

The exchange interactions in NiO and MnO have been investigated from first-principles calculations in a number of prior works~\cite{Zhang:2006, Fischer:2009, Archer:2011, Gopal:2017, Jacobsson:2013, Solovyev:1998, Logemann:2017, MacEnulty:2023}. These studies used different exchange-correlation functionals and corrective methods to describe the ground state, in particular, self-interaction correction (SIC) methods~\cite{Fischer:2009, Archer:2011}, hybrid functionals~\cite{Archer:2011}, and DFT+$U$ with either empirical or \textit{ab initio} $U$ values~\cite{Zhang:2006, Jacobsson:2013, Solovyev:1998, Gopal:2017, Logemann:2017, MacEnulty:2023}. These approaches differ in how they correct self-interaction errors, leading to variations in the resulting electronic structure, e.g. band gaps, magnetic moments, and charge densities. On top of these differing ground states, these studies employed various methods to compute $J_{ij}$ (e.g., MFT, $\Delta E$, or spin spiral), producing a spread of reported values with no clear consensus on either their absolute magnitudes or relative trends. Hence, comparisons of $J_{ij}$ are complicated due to the interplay of two factors: differences in the underlying ground-state description and differences in the computational approach used to determine the exchange parameters. Moreover, within the DFT+$U$ framework, the strong sensitivity of $J_{ij}$ to the value of $U$ further complicates comparisons between studies that used different $U$ values. Additionally, except for the study in Ref.~\citenum{Jacobsson:2013}, most research on NiO and MnO has focused on averaged isotropic exchange parameters $J_{ij}$, which neglects the splitting of nearest-neighbor $J_{ij}$ due to the rhombohedral distortion of the lattice. Moreover, no study has systematically compared MIPs from direct methods with those determined indirectly through LSWT fitting of magnon dispersion computed via dynamical spin susceptibility, such as using TDDFPT+$U$. Therefore, a comprehensive comparison of $J_{ij}$ values for transition-metal compounds, computed using both direct and indirect methods on top of an identical ground state and fully from first principles, is currently lacking.

In this study, we perform a systematic comparison of the isotropic exchange interaction parameters $J_{ij}$ and magnon dispersions in NiO and MnO using three first-principles methods. The first two are direct methods: the $\Delta E$ method based on FSMA~\cite{Whangbo:2003, Xiang:2011, Xiang:2013a, Sabani:2020}, and the IRM based on MFT~\cite{Liechtenstein:1987, Katsnelson:2000, Szilva:2023}, implemented using a plane-wave basis and wannierization~\cite{Korotin:2015, He:2021}. The $J_{ij}$ values obtained from both methods are then used to compute magnon dispersions via LSWT. The third approach is indirect: the TDDFPT+
$U$ method~\cite{Binci:2025}, based on the Liouville-Lanczos formalism~\cite{Gorni:2018}. This method computes the dynamical spin susceptibility from first principles, from which the magnon dispersion is obtained directly and then fitted using LSWT to extract the $J_{ij}$ parameters.
All three methods are applied consistently on top of the same DFT+$U$ ground state, with the Hubbard $U$ parameter computed \textit{ab initio} using DFPT~\cite{Timrov:2018, Timrov:2021, Binci:2023}. The rhombohedral distortions of NiO and MnO are accounted for through DFT+$U$ structural optimizations. These distortions lead to a splitting of the nearest-neighbor exchange interactions for parallel and antiparallel spin pairs due to magnetoelastic coupling. Our results show that both the TDDFPT+$U$ method and the LSWT approach using $J_{ij}$ values from the 
$\Delta E$ method yield magnon dispersions in good agreement with inelastic neutron scattering data. In particular, both approaches accurately reproduce the finite magnon energy at the M point in the Brillouin zone (BZ), especially for MnO. In contrast, the LSWT results based on $J_{ij}$ values from the MFT method exhibit less accurate agreement with experiments, which may be attributed to implementation-specific details of this approach~\cite{MartinezCarracedo:2023}.

\section{Computational details}
\label{sec:comput_details}

All calculations are performed using \QE\ (v7.2)~\cite{Giannozzi:2009, Giannozzi:2017, Giannozzi:2020}. Ground-state calculations are conducted with the \textsc{PW} code~\cite{Giannozzi:2009} using local spin-density approximation (LSDA) for the exchange-correlation functional including the Hubbard $U$ correction (LSDA+$U$)~\cite{Perdew:1992, Anisimov:1991}. We use norm-conserving scalar-relativistic pseudopotentials~\cite{Hamann:2013} from the \textsc{PseudoDojo} library~\cite{vanSetten:2018}. For all calculations (unless otherwise stated), the plane-wave expansion of the Kohn-Sham wavefunctions is carried out with an 80~Ry kinetic-energy cutoff, and a 320~Ry cutoff is used for the charge density and potentials. The BZ is sampled using a $\Gamma$-centered $12 \times 12 \times 12$ $\mathbf{k}$ points grid for the 4-atom rhombohedral unit cell, unless stated otherwise, and type II antiferromagnetic (AFII) ordering is used (see Fig.~\ref{fig:crystal_structure}). Spin-orbit coupling is neglected. Geometry optimization is performed using the Broyden-Fletcher-Goldfarb-Shanno (BFGS) algorithm~\cite{Fletcher:1987}, using a higher kinetic-energy cutoff of 90~Ry for the Kohn-Sham wavefunctions and 360~Ry for the charge density and potentials, with convergence criteria set to $10^{-8}$~Ry for total energy, $10^{-5}$~Ry/bohr for forces, and $0.01$~Kbar for pressure. All calculations are performed using the optimized LSDA+$U$ structural parameters. For NiO, the rhombohedral lattice parameter is $a = 5.03$~\AA\ and the rhombohedral angle is $\alpha = 33.65^\circ$, while for MnO, $a = 5.32$~\AA\ and $\alpha = 34.16^\circ$. The resulting magnetic moments and band gap values can be found in Ref.~\citenum{Binci:2025}.

The Hubbard $U$ parameters are computed using DFPT~\cite{Timrov:2018, Binci:2023} (see Eq.~(4) in the Supplemental Material (SM)~\cite{SupplementalMaterial}) as implemented in the \textsc{HP} code~\cite{Timrov:2022}, with L\"owdin-orthogonalized atomic orbitals for Hubbard projectors~\cite{Lowdin:1950}. We employ uniform $\Gamma$-centered $\mathbf{k}$ and $\mathbf{q}$ points grids of size $8 \times 8 \times 8$ and $4 \times 4 \times 4$, respectively, for the 4-atom AFII unit cell, achieving an accuracy of approximately 0.01~eV for the Hubbard parameters. The $U$ parameters are computed iteratively in a self-consistent manner, as outlined in Ref.~\cite{Timrov:2021}, and include Hubbard forces and stresses in LSDA+$U$ structural optimizations~\cite{Timrov:2020}. The final computed values for $U$ are 6.26~eV for the Ni($3d$) and 4.29~eV for the Mn($3d$) states in NiO and MnO, respectively.

The magnon energies are computed using TDDFPT+$U$~\cite{Binci:2025}, as implemented in a \textsc{turboMagnon} code~\cite{Gorni:2022}. This method is based on computing the dynamical spin susceptibility using the Liouville-Lanczos approach~\cite{Gorni:2018} (see Eqs.~(21) and (23) in the SM~\cite{SupplementalMaterial}). We use adiabatic LSDA+$U$ (ALSDA+$U$). The 4-atom AFII unit cell is used and the BZ is sampled with a $\Gamma$-centered $12 \times 12 \times 12$ $\mathbf{k}$ points grid. The calculations employ the pseudo-Hermitian flavor of the Lanczos recursive algorithm~\cite{Gorni:2023, Delugas:2023}, which includes an extrapolation technique for the Lanczos coefficients~\cite{Rocca:2008}. A total of 8000 Lanczos iterations are performed to achieve convergence in the TDDFPT+$U$ calculations. The TDDFPT+$U$ exchange parameters are derived by fitting the TDDFPT+$U$ magnon dispersions using LSWT (see Eqs.~(7)--(9) in the SM~\cite{SupplementalMaterial}).

The exchange interaction parameters are computed using MFT (see Eqs.~(18) in the SM~\cite{SupplementalMaterial}) as implemented in the TB2J code (v0.7.7.2)~\cite{He:2021}. Calculations are performed with the 4-atom AFII unit cell. MLWFs are generated using the Wannier90 code (v3.1.0)~\cite{Pizzi:2020}, based on the LSDA+$U$ ground state. Specifically, we generate MLWFs for Mn($3d$), Mn($4s$), and O($2p$) states in MnO, as well as for Ni($3d$), Ni($4s$), and O($2p$) states in NiO, resulting in a total of 18~MLWFs out of 45 considered Kohn-Sham states for each material. We confirm that the Wannier centers for the Mn($3d$) and Ni($3d$) states are accurately centered on the atoms. For the ground-state calculations, the MLWFs generation, and the calculation of $J_{ij}$, the BZ is sampled using a $\Gamma$-centered $18 \times 18 \times 18$ $\mathbf{k}$ points grid. This sampling ensures that the calculation of the $J_{ij}$ parameters converges with an accuracy of 0.01~meV.

To compute the exchange interaction parameters using the $\Delta E$ method, we have developed a workflow within the AiiDA package~\cite{Huber:2020}, which will be described in detail elsewhere. We start with the AFII ground state in the LSDA+$U$ framework for the 4-atom unit cell, and then construct supercells of size $4 \times 4 \times 4$. The BZ is sampled using a $\Gamma$-centered $4 \times 4 \times 4$ $\mathbf{k}$ points grid for NiO and a $3 \times 3 \times 3$ $\mathbf{k}$ points grid for MnO. The plane-wave expansion of the Kohn-Sham wavefunctions and charge density employs kinetic-energy cutoffs of 80~Ry and 320~Ry for NiO, while for MnO we used 100~Ry and 400~Ry, respectively. This ensures the convergence of the $J_{ij}$ parameters with an accuracy of 0.01~meV for both materials. In these supercells, we initialize the original AFII configuration and then consider a target pair of spins in the following configurations: $\uparrow \uparrow$ (state~1), $\uparrow \downarrow$ (state~2), $\downarrow \uparrow$ (state~3), and $\downarrow \downarrow$ (state~4). For each configuration, we compute the total energy and then evaluate the exchange interaction parameters [see Eq.~(13) in the SM~\cite{SupplementalMaterial}]. This procedure allows us to determine the exchange parameters $J_{ij}$ depending on the selected target pair of spins in the supercell.

The magnon dispersions based on LSWT, using the exchange interaction parameters reported in Table~\ref{tab:exchange_param_from_DFT}, are computed with the SpinW code (v3.1)~\cite{Toth:2015}. The fitting of all magnon dispersions in Fig.~\ref{fig:dispersions} is performed using the minimal model [see Eq.~\eqref{eq:heisenberg_hamiltonian}] and the analytic LSWT expressions [see Eqs.~(7)--(9) in the SM~\cite{SupplementalMaterial}].

The data used to produce the results of this work is available in the Materials Cloud Archive~\cite{MaterialsCloudArchive2025}.

\section{RESULTS} 

\subsection{Magnetic order and Heisenberg Hamiltonians for NiO and MnO}

\begin{figure}[h!]
\centering
\includegraphics[width=0.38\textwidth]{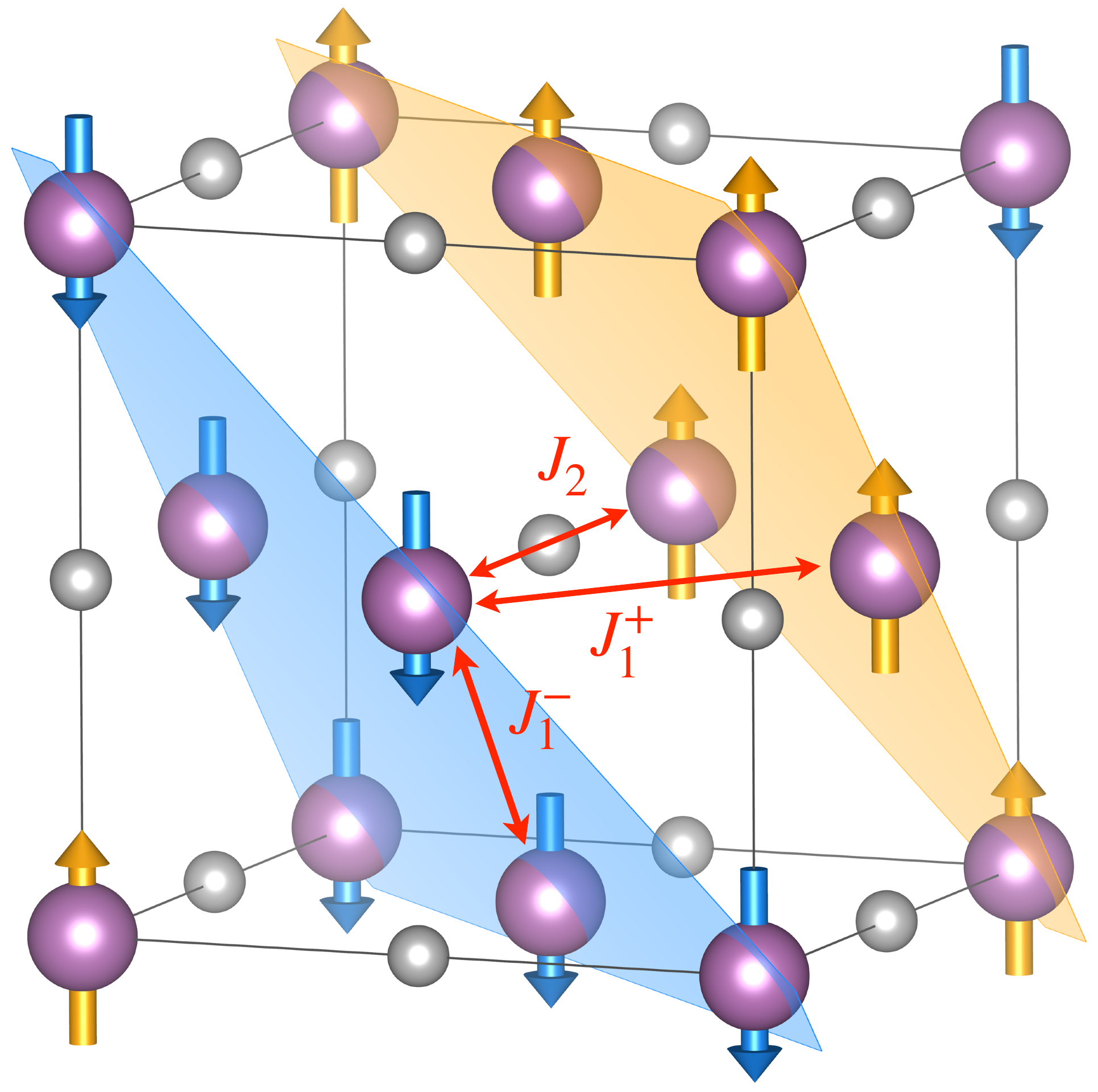}
\caption{Face-centered cubic conventional unit cell of NiO and MnO with the AFII magnetic order. Transition-metal elements (Ni in NiO, or Mn in MnO) and O atoms are shown in purple and grey, respectively. Yellow and blue vertical thick arrows centered on atoms indicate the direction of the spin. The ferromagnetic (111) planes are depicted in transparent blue and yellow color. The Heisenberg exchange interaction parameters $J_1^+$, $J_1^-$, and $J_2$ are highlighted for selected atoms, with thin red arrows showing the atom pairs involved in these interactions. Rendered using \textsc{VESTA}~\cite{Momma:2008}.}
\label{fig:crystal_structure}
\end{figure}

NiO and MnO are prototypical insulating transition-metal monoxides, widely used to benchmark and compare computational methods for describing structural, electronic, magnetic, and spectral properties. At high temperatures, both materials crystallize in the rocksalt-type structure and adopt a paramagnetic phase. Upon cooling below their respective Néel temperatures, 524~K for NiO~\cite{Srinivasan:1984} and 120~K for MnO~\cite{Blech:1964}, they undergo a transition to a type II antiferromagnetic (AFM) phase. This magnetic ordering is accompanied by a small rhombohedral distortion along the [111] direction of the face-centered cubic (fcc) lattice. In the AFM phase, ferromagnetic (111) planes are stacked antiferromagnetically along the [111] direction, as illustrated in Fig.~\ref{fig:crystal_structure}.

\begin{table*}[t!]
    \renewcommand{\arraystretch}{1.}
    \setlength\tabcolsep{0.1in}
    \centering   
    \begin{tabular}{clccccccccccc}
        \hline\hline  
        Material             & Method     & $J_1^+$ & $J_1^-$ & $J_2$ & $J_{3}^+$ & $J_{3}^-$ & $J_4$ & $J_5$ & \\ 
        \hline
        \multirow{2}{*}{NiO} & MFT        & $-$0.12 & $-$0.16 & 13.88 &    0.01   & 0.00      & 0.09  & 0.03  &  \\              
                             & $\Delta E$ & $-$1.21 & $-$0.95 & 12.57 & $-$0.04   & $-$0.04      & 0.49  & 0.01  &  \\
        \hline
        \multirow{2}{*}{MnO} 
                             & MFT        &    1.23 &    0.83 &  1.03 &    0.00   & 0.01      & 0.03  & 0.02  &  \\ 
                             & $\Delta E$ &    0.47 &    0.36 &  0.59 &    0.00   & 0.00      & 0.05  & 0.01  &  \\ 
        \hline\hline   
    \end{tabular}
    \caption{Magnetic exchange interaction parameters (in meV) for NiO and MnO using the convention for the Heisenberg Hamiltonian in Eq.~\eqref{eq:heisenberg_hamiltonian_0}, as obtained using MFT and $\Delta E$.}
    \label{tab:exchange_param_from_DFT}
\end{table*}

To describe spin-wave excitations in NiO and MnO using direct methods, we postulate a Heisenberg Hamiltonian that includes only isotropic (but, in principle, long range) two-spin exchange interactions:
\begin{equation}
H_\mathrm{spin} = \sum_{i,j} J_{ij} {\bm{S}}_i \cdot {\bm{S}}_j ,
\label{eq:heisenberg_hamiltonian_0}
\end{equation}
where $i$ and $j$ label magnetic atomic sites, $J_{ij}$ denotes the isotropic exchange interaction between spins at sites $i$ and $j$, while ${\bm{S}}_i$ and ${\bm{S}}_j$ are the corresponding spin operators. In this work, we adopt the convention that the spin operators are not normalized to unity, but instead satisfy $|\bm{S}_i| = |\bm{S}_j| = S$, where $S$ is the spin quantum number. The Hamiltonian in Eq.~\eqref{eq:heisenberg_hamiltonian_0} neglects additional magnetic interactions such as the Dzyaloshinskii–Moriya, single-ion anisotropy, and higher-order exchange terms. Furthermore, we note that each interaction pair is counted twice in the summation, i.e., both $J_{ij}$ and $J_{ji}$ are included. 

Due to the small rhombohedral distortion of the lattice at low temperatures, certain exchange interactions become inequivalent for parallel and antiparallel spin orientations. In particular, the nearest-neighbor exchange interaction $J_1$ splits into two distinct parameters, $J_1^{-}$ and $J_1^{+}$ -- a notation introduced by Lines and Jones~\cite{Lines:1965} -- to distinguish between interactions involving parallel and antiparallel spin pairs, respectively. This splitting originates from magnetoelastic coupling induced by the distortion. In the absence of such structural distortions, these parameters are equal, i.e., $J_1^{-} = J_1^{+}$. In contrast, the next-nearest-neighbor exchange interaction $J_2$ occurs only between antiparallel spin pairs, and thus a single parameter is sufficient to describe it. These exchange interactions are illustrated in Fig.~\ref{fig:crystal_structure}; interactions beyond the second-neighbor shell are not shown. A detailed analysis of the physical origin and behavior of exchange interactions in these systems can be found, for example, in Refs.~\citenum{Fischer:2009, Archer:2011}.

\subsection{First-principles calculations of exchange parameters using the direct MFT and $\Delta E$ methods}

As discussed, we first compute magnetic exchange parameters using two direct first-principles approaches: MFT and $\Delta E$. Both methods are applied consistently on top of the same DFT+$U$ ground state, using identical Hubbard $U$ values obtained from DFPT. The inclusion of $U$ is essential for NiO and MnO, as it corrects self-interaction errors of (semi-)local functionals and thus properly localizes the $d$ electrons. This consistent setup contrasts with previous studies~\cite{Archer:2011, Gopal:2017, Jacobsson:2013, Solovyev:1998}, where a single method was used to compute exchange parameters but with varying ground-state descriptions. By fixing the ground state using DFT+$U$, we investigate the accuracy of each method (MFT and $\Delta E$) in predicting the MIPs, similar to the strategy in Ref.~\citenum{Fischer:2009} (which employed the SIC method for the ground-state description). Moreover, using first-principles $U$ values removes ambiguity in the choice of $U$, avoiding additional spread in the computed MIPs.

Each of the methods considered yields individual exchange parameters $J$, which must be computed for as many neighbor shells as needed to ensure convergence of the magnon dispersions. For both NiO and MnO, we find that including exchange interactions up to the fifth nearest neighbor shell is necessary to achieve stable (i.e., positive-definite) magnon dispersions with convergence within a few meV. The corresponding values are listed in Table~\ref{tab:exchange_param_from_DFT}. While most previous studies~\cite{Zhang:2006, Fischer:2009, Archer:2011, Gopal:2017, Jacobsson:2013, Solovyev:1998, Logemann:2017, MacEnulty:2023} limit the analysis to only $J_1$ and $J_2$ (often neglecting the splitting of $J_1$), our results highlight the importance of computing further-neighbor couplings to obtain stable and quantitatively accurate magnon dispersions. We also find that the splitting between parallel and antiparallel spin interactions occurs not only for the first nearest-neighbor shell ($J_1$) but also for the third nearest-neighbor shell ($J_3$) when the MFT method is used. All reported $J$ values are expressed using spin quantum numbers $S=1$ for Ni$^{2+}$ and $S=5/2$ for Mn$^{2+}$ (see Eq.~\eqref{eq:heisenberg_hamiltonian_0}). Alternative definitions, such as setting $S$ equal to half the calculated magnetic moment, exist in the literature, but suffer from ambiguity in the definition of the magnetic moment. Therefore, the choice of $S$ must be taken into account when comparing the exchange parameters across different computational studies or with experimentally fitted values, as will be discussed below.

\begin{figure*}[t!]
\centering
\includegraphics[width=0.48\textwidth]{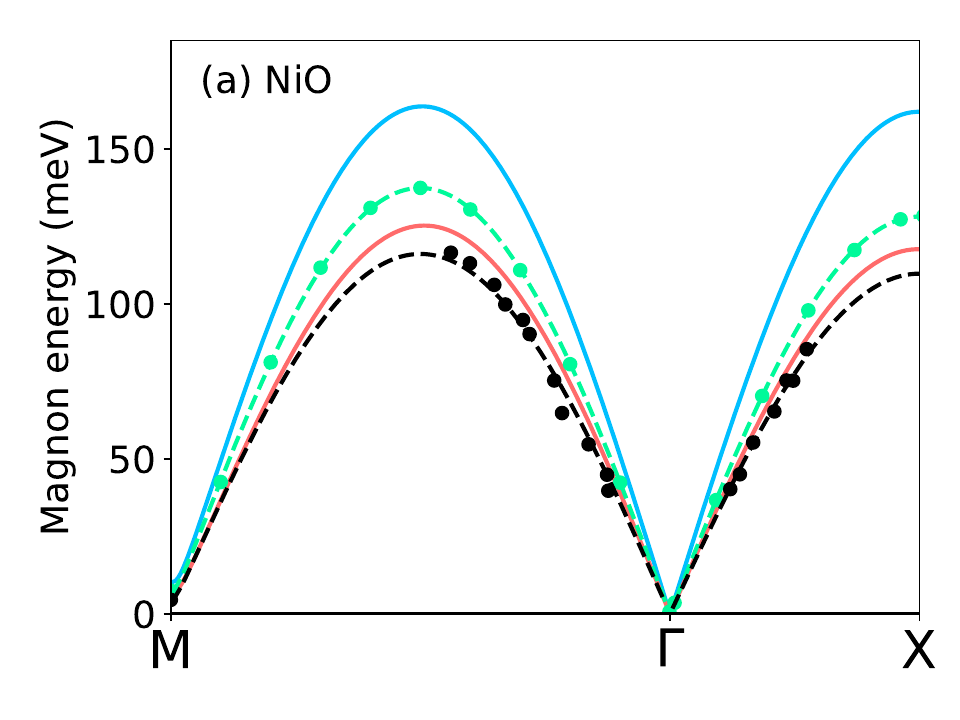}
\includegraphics[width=0.48\textwidth]{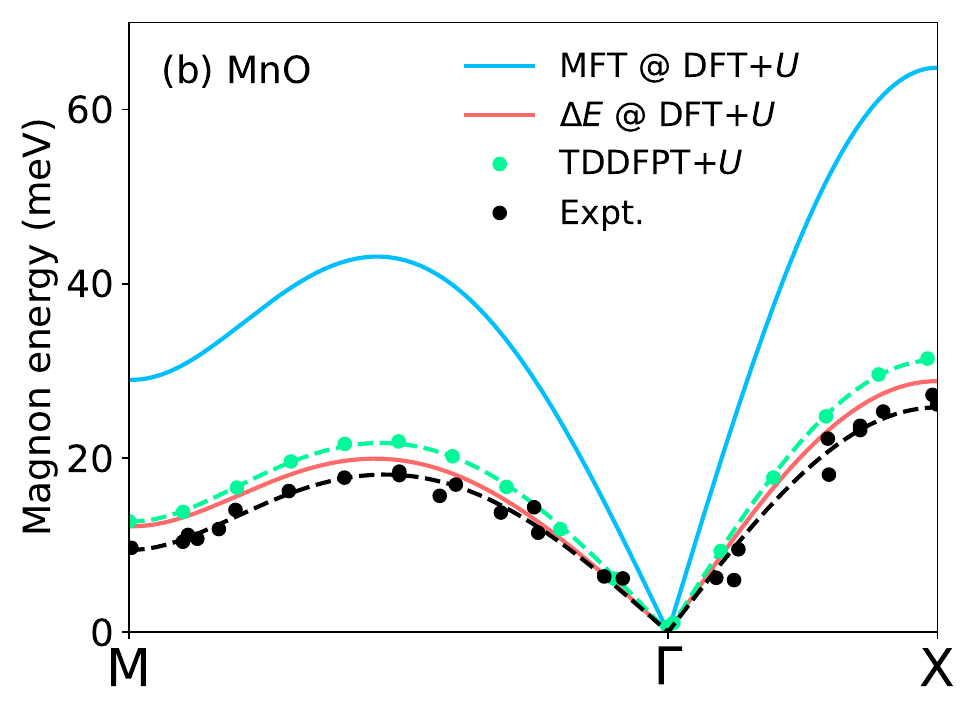}
\caption{
Magnon dispersions for (a) NiO and (b) MnO computed using LSWT with Heisenberg exchange parameters from MFT and $\Delta E$ are shown as blue and red solid lines, respectively (see Table~\ref{tab:exchange_param_from_DFT} and Eq.~\eqref{eq:heisenberg_hamiltonian_0}). The TDDFPT+$U$ magnons dispersions are computed directly from the dynamical spin susceptibility and are shown as green dots. Experimental magnon dispersions, shown as black dots, are taken from inelastic neutron scattering measurements reported in Refs.~\citenum{Hutchings:1972, Pepy:1974}. The green and black dashed lines represent LSWT fits to the TDDFPT+$U$ and experimental data, respectively, based on the effective Heisenberg Hamiltonian of Eq.~\eqref{eq:heisenberg_hamiltonian}. All theoretical results are based on the same DFT+$U$ ground state.}
\label{fig:dispersions}
\end{figure*} 

For NiO, both MFT and $\Delta E$ yield similar values for the dominant antiferromagnetic exchange interaction $J_2$, with MFT predicting a value approximately 10\% higher than that from $\Delta E$. Consistently with previous computational studies, $J_2$ is the largest exchange parameter. In contrast, $J_1^+$ and $J_1^-$ are 1–2 orders of magnitude smaller and have opposite sign to $J_2$. However, their values differ significantly between methods: MFT predicts $|J_1^+| < |J_1^-|$, which favors the stabilization of the observed antiferromagnetic spin structure. In contrast, $\Delta E$ gives $|J_1^+| > |J_1^-|$, which destabilizes the magnon spectrum unless further-neighbor interactions are included. For the third to fifth neighbor interactions, both methods predict very small values. MFT yields negligible values, with only $J_4$ being about 1.5–2 times smaller than $J_1^+$ and $J_1^-$. A similar trend is observed in $\Delta E$, where only $J_4$ is sizable. Although these couplings are minor compared to $J_2$, they are essential in the $\Delta E$ approach to stabilize magnons, while their impact in MFT is minimal (see Fig.~S1(a) in the SM~\cite{SupplementalMaterial}).

For MnO, the dominant exchange interactions are $J_1^+$, $J_1^-$, and $J_2$, all having similar magnitudes and the same (positive) sign within each method, unlike in NiO. However, MFT and $\Delta E$ predict different trends: MFT gives $|J_1^+| > |J_2| > |J_1^-|$, while $\Delta E$ yields $|J_2| > |J_1^+| > |J_1^-|$. The inclusion of these three parameters alone is sufficient to stabilize magnons in both approaches. However, their absolute values are consistently larger in MFT than in $\Delta E$ by roughly a factor of two, leading to similarly scaled differences in magnon energies [see Fig.~S1(b) in the SM~\cite{SupplementalMaterial}]. Third to fifth nearest-neighbor interactions are 1–2 orders of magnitude smaller than $J_1^+$, $J_1^-$, and $J_2$. Despite their small magnitude, including these further-neighbor interactions is necessary to achieve convergence of the magnon dispersion. Omitting them leads to energy deviations of up to 4~meV for certain magnon momenta, corresponding to errors of about 15\%. Thus, while these interactions are not essential for magnon stability (as they are in NiO) they are important for the quantitative comparison with the experimental data.

\subsection{Magnon dispersions}
\label{sec:magnons}

We now compare the three computational approaches considered, to predict magnon dispersions in NiO and MnO. The first two are based on LSWT using Heisenberg exchange parameters $J$ obtained from MFT and $\Delta E$, as discussed in the previous section. The third approach computes magnon dispersions directly via TDDFPT+$U$ by evaluating the dynamical spin susceptibility~\cite{Binci:2025}. A brief overview of each method is provided in Sec.~S3 of the SM~\cite{SupplementalMaterial}.  All methods are applied consistently on top of the same DFT+$U$ ground state, enabling a direct comparison focused on the physical approximations inherent to each method in describing magnetic excitations. The computed magnon spectra are benchmarked against inelastic neutron scattering data from Refs.~\citenum{Hutchings:1972, Pepy:1974}.

The primitive magnetic unit cell of NiO and MnO contains two antiferromagnetically aligned magnetic ions due to the AFII spin configuration. In the absence of magnetic anisotropy, this leads to two degenerate magnon modes. However, weak magnetocrystalline anisotropies (arising from spin-orbit coupling) lift this degeneracy near the $\Gamma$ point, resulting in a small magnon gap at $\Gamma$ of less than 5~meV in both materials~\cite{Lines:1965, Hutchings:1972, Pepy:1974}. Although this gap is small in NiO and more pronounced in MnO, relative to the maximum magnon energies in each material, we follow common practice in the literature and neglect magnetocrystalline anisotropy, treating the magnon gap at $\Gamma$ as zero. This simplification is justified because our primary aim is to compare the accuracy of different computational methods in predicting magnon dispersions across the full BZ, rather than focusing on fine features near the $\Gamma$ point.

Figure~\ref{fig:dispersions} shows the magnon dispersions computed using the three methods along the $\Gamma$–M and $\Gamma$–X directions in the BZ for rhombohedrally distorted NiO and MnO. TDDFPT+$U$ and LSWT based on $\Delta E$-derived exchange parameters both reproduce the experimental data well, capturing the overall trends and curvatures of the magnon bands. The maximum magnon energies are overestimated on average by about 15~meV (14\%) for NiO and 3~meV (17\%) for MnO. Between the two methods, LSWT with $\Delta E$-derived parameters shows better agreement with experiment than TDDFPT+$U$. This good agreement of both methods with experiments is remarkable given that both methods are fully first-principles and rely exclusively on \textit{ab initio} Hubbard $U$ values, without any empirical fitting. In contrast, LSWT using MFT-derived $J$ values, while qualitatively reproducing the general features of the dispersion, significantly overestimates the magnon energies. The deviations reach 46~meV (40\%) for NiO and 25~meV (139\%) for MnO. Such a large deviation of magnon energies from MFT, especially for MnO, is quite striking when compared to the other two methods.

A key feature of the magnon dispersion in NiO and MnO is the finite magnon energy at the M point in the BZ. This feature arises only when the rhombohedral lattice distortion is taken into account~\cite{Jacobsson:2013, Binci:2025}, which splits the nearest-neighbor (and some further-neighbor) exchange interaction into two distinct parameters, $J_1^+$ and $J_1^-$. If the distortion is neglected, then $J_1^+ = J_1^-$, and the magnon energy at M becomes zero. In our calculations, which include the rhombohedral distortion, all three methods correctly predict a finite magnon energy at M. However, the values obtained from TDDFPT+$U$ and LSWT based on $\Delta E$-derived exchange parameters agree much better with experiment than those from LSWT based on MFT-derived parameters. This magnon energy at M is relatively small in NiO compared to its maximum magnon energy, but significantly larger in MnO. This difference reflects the relative magnitudes of the exchange interactions: in NiO, $J_2$ dominates and exceeds $J_1^+$ and $J_1^-$ by 1–2 orders of magnitude, whereas in MnO, all three parameters are of similar magnitude (see Table~\ref{tab:exchange_param_from_DFT}).

\begin{figure*}[t!]
\centering
\includegraphics[width=0.99\textwidth]{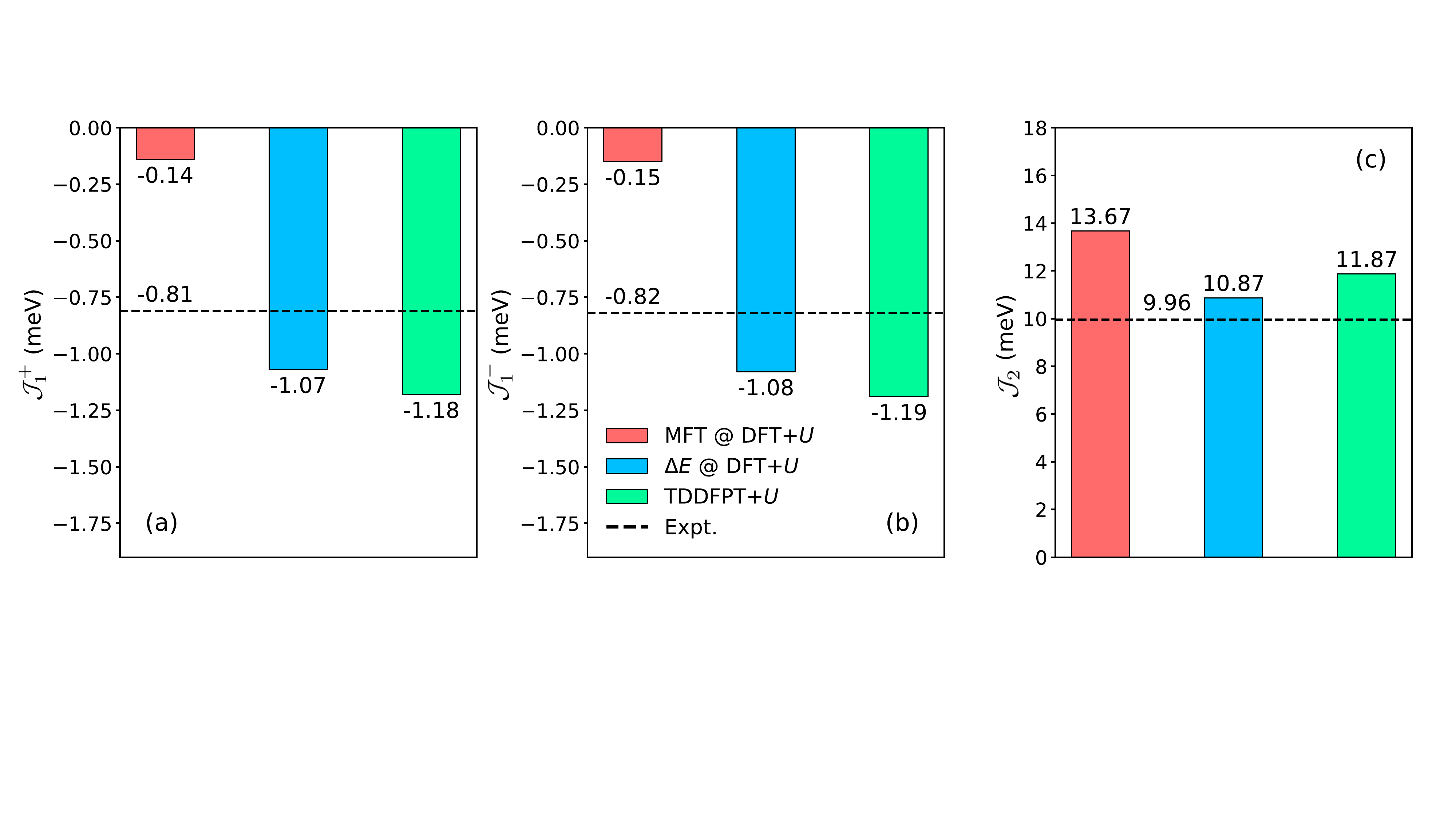}
\caption{Effective exchange parameters for NiO within the minimal Heisenberg model [Eq.~\eqref{eq:heisenberg_hamiltonian}]: (a) $\mathcal{J}_1^+$, (b) $\mathcal{J}_1^-$, and (c) $\mathcal{J}_2$, all in meV, extracted by fitting magnon dispersions 
using the analytical LSWT expression (see Sec.~S3 in the SM~\cite{SupplementalMaterial}) for four data sets:
$(i)$~LSWT dispersions with MFT-derived $J_1$--$J_5$ values,
$(ii)$~LSWT dispersions with $\Delta E$-derived $J_1$--$J_5$ values,
$(iii)$~TDDFPT+$U$ magnon dispersions from the dynamical spin susceptibility,
$(iv)$~Experimental data from Ref.~\citenum{Hutchings:1972}.
Theoretical results are shown as histograms, while experimental values are indicated by horizontal dashed lines. All theoretical results are based on the same DFT+$U$ ground state.}
\label{fig:NiO}
\end{figure*} 
\begin{figure*}[t!]
\centering
\includegraphics[width=0.99\textwidth]{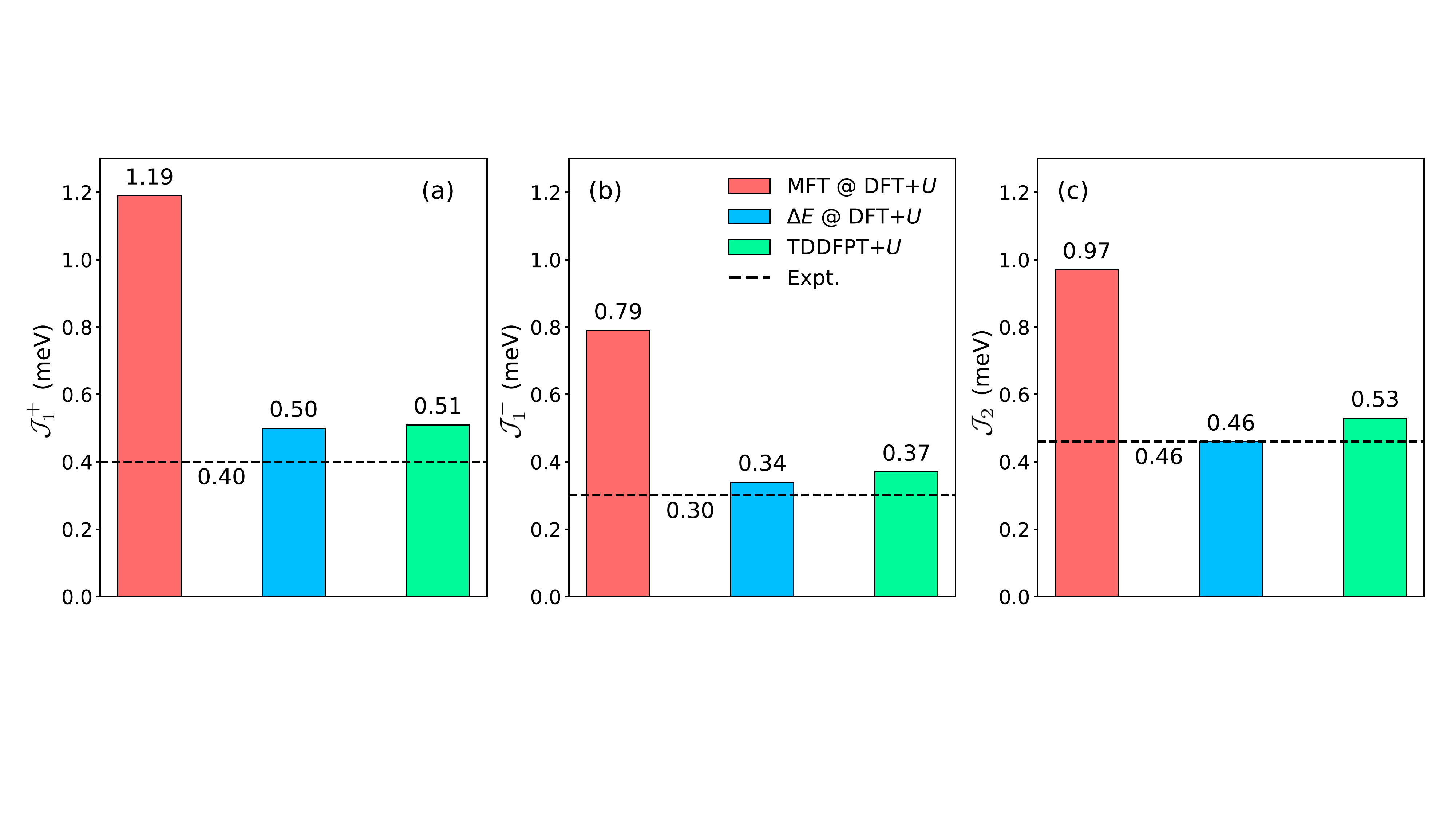}
\caption{Effective exchange parameters for MnO within the minimal Heisenberg model [Eq.~\eqref{eq:heisenberg_hamiltonian}]: (a) $\mathcal{J}_1^+$, (b) $\mathcal{J}_1^-$, and (c) $\mathcal{J}_2$, all in meV, extracted by fitting magnon dispersions 
using the analytical LSWT expression (see Sec.~S3 in the SM~\cite{SupplementalMaterial}) for four data sets:
$(i)$~LSWT dispersions with MFT-derived $J_1$--$J_5$ values,
$(ii)$~LSWT dispersions with $\Delta E$-derived $J_1$--$J_5$ values,
$(iii)$~TDDFPT+$U$ magnon dispersions from the dynamical spin susceptibility,
$(iv)$~Experimental data from Ref.~\citenum{Pepy:1974}.
Theoretical results are shown as histograms, while experimental values are indicated by horizontal dashed lines. All theoretical results are based on the same DFT+$U$ ground state.}
\label{fig:MnO}
\end{figure*} 

\subsection{Fitting effective exchange parameters of the minimal Heisenberg model}

Inelastic neutron scattering data for NiO and MnO~\cite{Hutchings:1972, Pepy:1974} are typically analyzed using LSWT based on the Heisenberg Hamiltonian with a fixed number of MIPs. These MIPs are fitted to the experimental data and then used to characterize the magnetic exchange interactions in these materials. The number of MIPs is chosen to ensure a good fit to the experimental magnon dispersion, while keeping the model as simple as possible. As we showed previously, from a computational standpoint, accurate convergence of the magnon dispersion in NiO and MnO requires including exchange interactions up to the fifth nearest neighbor shell (see Table~\ref{tab:exchange_param_from_DFT}). However, fitting experimental data with such a large number of parameters is challenging due to limited resolution and uncertainties in the measurements, and because distant interactions are small in magnitude. Therefore, experimental studies typically use a reduced model, fitting only the nearest and next-nearest neighbor interactions, as described by the simplified Heisenberg Hamiltonian~\cite{Hutchings:1972, Pepy:1974}:
\begin{eqnarray}  
    H_\mathrm{spin} & = & \sum_{i,j}^\mathrm{n.n.p} \mathcal{J}_1^{-} \,{\bm{S}}_i \cdot {\bm{S}}_j + \sum^\mathrm{n.n.a}_{i,j} \mathcal{J}_1^{+}\,{\bm{S}}_i\cdot{\bm{S}}_{j} \nonumber \\
    & & + \sum_{i,j}^\mathrm{n.n.n.} \mathcal{J}_2\,{\bm{S}}_i\cdot{\bm{S}}_{j} \,.
    \label{eq:heisenberg_hamiltonian}
\end{eqnarray}
Here, $\mathcal{J}_1^{-}$ and $\mathcal{J}_1^{+}$ are \textit{effective} nearest-neighbor exchange parameters for the parallel (n.n.p) and antiparallel (n.n.a) spins, respectively, that differ due to rhombohedral lattice distortions, while $\mathcal{J}_2$ is the \textit{effective} next-nearest-neighbor (n.n.n.) exchange parameter. These effective parameters differ from the exchange parameters $J_1^{-}$, $J_1^{+}$, and $J_2$ in Eq.~\eqref{eq:heisenberg_hamiltonian_0} because the former ones are renormalized to account for the absence of third- to fifth-nearest-neighbor interactions in the simplified model of Eq.~\eqref{eq:heisenberg_hamiltonian}. That is, the effect of these neglected interactions is effectively absorbed into the first two shells. To clearly differentiate these fitting parameters $\mathcal{J}$ from the exchange couplings $J$ computed using MFT or $\Delta E$, we use a different font style in our notation.

To fit all magnon dispersion curves in Fig.~\ref{fig:dispersions} using the minimal model of Eq.~\eqref{eq:heisenberg_hamiltonian} and the respective analytical LSWT expression (see Sec.~S3 in the SM~\cite{SupplementalMaterial}), we employ a non-linear least-squares procedure. Importantly, we perform the fitting independently rather than relying on the fitted values reported in Refs.~\citenum{Hutchings:1972, Pepy:1974}, for two main reasons. First, the original works included a single-ion anisotropy term arising from magnetocrystalline effects, which we neglect in Eq.~\eqref{eq:heisenberg_hamiltonian}. We verified that this term only slightly renormalizes $\mathcal{J}_1^{-}$, $\mathcal{J}_1^{+}$, and $\mathcal{J}_2$ (at most by a few percent), and therefore we omit it to ensure a consistent comparison focused solely on the first- and second-nearest neighbor exchange interactions. Second, we adopt spin quantum numbers $S = 1$ for Ni$^{2+}$ and $S = 5/2$ for Mn$^{2+}$, while Ref.~\citenum{Pepy:1974} used $S = 2.445$ for Mn$^{2+}$. Although this is a minor difference, we use nominal spin values to ensure maximal consistency with our computational models and to enable an accurate comparison of the fitted exchange parameters. Finally, for NiO, the magnon energy at the M point in Fig.~\ref{fig:dispersions} is essential for a reliable fit due to the limited number of experimental points along the $\Gamma$--M path in the BZ. We use the value 4.54~meV, consistent with Ref.~\citenum{Hutchings:1972}, which was based on infrared measurements from Refs.~\citenum{Kondoh:1960, Sievers:1963}. 

The fitting curves to the experimental and TDDFPT+$U$ magnon dispersions are shown in Fig.~\ref{fig:dispersions}, while the corresponding fits of the LSWT-derived magnon dispersions using the $J_1$--$J_5$ exchange parameters from MFT and $\Delta E$ (listed in Table~\ref{tab:exchange_param_from_DFT}) are presented in Sec.~S2 of the SM~\cite{SupplementalMaterial}. As seen in Fig.~\ref{fig:dispersions}, the TDDFPT+$U$ data are fitted with high accuracy. The fit to the experimental data is also satisfactory, considering the substantial spread in the measurements caused by the limited resolution of inelastic neutron scattering. Similarly, the fitted curves show excellent agreement with the LSWT dispersions based on the MFT and $\Delta E$ parameters (see Sec.~S2 of the SM~\cite{SupplementalMaterial}). This confirms that the minimal Heisenberg model in Eq.~\eqref{eq:heisenberg_hamiltonian} is sufficient to reproduce all the features of the magnon dispersions in NiO and MnO with high accuracy. In the following, we compare the fitted exchange parameters $\mathcal{J}_1^{-}$, $\mathcal{J}_1^{+}$, and $\mathcal{J}_2$ with those computed from first principles using the MFT and $\Delta E$ methods, namely $J_1^{-}$, $J_1^{+}$, and $J_2$.

Figure~\ref{fig:NiO} summarizes the effective exchange parameters $\mathcal{J}_1^{-}$, $\mathcal{J}_1^{+}$, and $\mathcal{J}_2$ for NiO, obtained by fitting the magnon dispersions from TDDFPT+$U$ and LSWT-derived ones using the $J_1$--$J_5$ values from MFT and $\Delta E$, and compare them to those fitted to experimental data~\cite{Hutchings:1972, Pepy:1974}. The MFT-fitted parameters deviate most from experiment, while the $\Delta E$-fitted parameters are in closest agreement. TDDFPT+$U$-fitted parameters yield intermediate accuracy, slightly less accurate than the $\Delta E$-fitted values. Despite differences in accuracy, all methods consistently reproduce the trend $|\mathcal{J}_1^+| < |\mathcal{J}_1^-| \ll |\mathcal{J}_2|$. Notably, the trend $|\mathcal{J}_1^+| < |\mathcal{J}_1^-|$ is crucial for ensuring positive-definiteness of the magnon spectrum in the minimal Heisenberg model~\eqref{eq:heisenberg_hamiltonian}. We recall that this trend is reversed in the raw $\Delta E$ data for NiO (see Table~\ref{tab:exchange_param_from_DFT}), leading to a magnon instability near the M point [see Fig.~S1(a) in the SM~\cite{SupplementalMaterial}]. The comparison of the fitted effective ($\mathcal{J}$) and computed ($J$) parameters reveals systematic renormalizations: The average deviation $\Delta(\mathcal{J}_1^+ - J_1^+)$ is 14\% and $\Delta(\mathcal{J}_1^- - J_1^-)$ is 10\% for both MFT and $\Delta E$, while $\Delta(\mathcal{J}_2 - J_2)$ is 2\% for MFT and 14\% for $\Delta E$.  These differences underscore how excluding third- to fifth-nearest-neighbor terms and absorbing their effects into renormalized $\mathcal{J}_1^{-}$, $\mathcal{J}_1^{+}$, and $\mathcal{J}_2$ in the minimal model influences the resulting magnon dispersion in NiO. 

Figure~\ref{fig:MnO} shows the effective fitted exchange parameters $\mathcal{J}_1^{-}$, $\mathcal{J}_1^{+}$, and $\mathcal{J}_2$ for MnO. As in NiO, MFT-fitted values deviate most from experiment, $\Delta E$-fitted values show the best agreement, and TDDFPT+$U$ yields intermediate accuracy. However, unlike NiO, the fitted parameter trends differ across methods: MFT and $\Delta E$ give $|\mathcal{J}_1^+| > |\mathcal{J}_2| > |\mathcal{J}_1^-|$, while TDDFPT+$U$ gives $|\mathcal{J}_2| > |\mathcal{J}_1^+| > |\mathcal{J}_1^-|$. Only the TDDFPT+$U$ trend matches the experimental trend. Moreover, the MFT-fitted trend matches the trend for computed $J$ values using MFT, while the $\Delta E$-fitted trend differs from its corresponding trend for computed $J$ values using $\Delta E$. Unlike in NiO, these variations in trends for MnO do not affect magnon stability. The comparison of effective ($\mathcal{J}$) and computed ($J$) parameters  reveals renormalization effects in MnO that differ in magnitude from those observed in NiO. More specifically, for MnO, the average $\Delta(\mathcal{J}_1^+ - J_1^+)$ is 5\% for MFT and $\Delta E$, which is 3 times smaller than in NiO; the average $\Delta(\mathcal{J}_1^- - J_1^-)$ is also 5\% for both methods, which is 2 times smaller than in NiO; $\Delta(\mathcal{J}_2 - J_2)$ is 6\% for MFT and 22\% for $\Delta E$, which is $2-3$ times larger than in NiO. These differences highlight the strong material dependence of the exchange parameter renormalization in the minimal Heisenberg model.

It is useful to comment on the trend observed in the renormalization of exchange parameters when the magnon dispersions derived from the full set of \textit{ab initio} couplings are fitted using the minimal Heisenberg model. For both NiO and MnO, the effective fitted parameters are generally smaller than the computed ones, which reflects how the longer-range interactions (when omitted explicitly) are effectively absorbed into the nearest- and next-nearest-neighbour terms. Their signs and relative magnitudes influence the curvature and overall dispersion in such a way that the minimal model reproduces the correct magnon energies only by adjusting (renormalizing) the retained couplings. Whether similar trends hold more broadly remains an open question and will require further systematic studies across different classes of magnetic materials.

\section{DISCUSSION}
\label{sec:discussion}

The findings of this study naturally raise important questions: Why do LSWT with $\Delta E$-derived exchange parameters and TDDFPT+$U$ give the best match with experiment? And why does MFT perform worst? In the following, we briefly analyze the fundamental differences between these approaches to shed light on these points.

The key features of TDDFPT+$U$ can be summarized as follows:
$(i)$~Direct calculation of magnon dispersions: TDDFPT+$U$ computes magnon spectra without relying on a low-energy model, such as the Heisenberg Hamiltonian in combination with LSWT. It includes magnetic pairwise exchange interactions to infinite order; i.e., inherently accounting for all neighbor interactions (e.g., $J_3$, $J_4$, $J_5$, and beyond) without any truncation.
$(ii)$~Fully dynamical formalism: Unlike static methods such as MFT and $\Delta E$, TDDFPT+$U$ operates in the frequency domain by solving time-dependent linear-response equations. This allows direct access to the dynamical spin susceptibility, including Hubbard $U$ corrections~\cite{Binci:2025}, thereby extending the DFT+$U$ framework to excited states. In contrast, dynamics in MFT and $\Delta E$ enters only through LSWT applied to a static spin Hamiltonian.
$(iii)$~Coupled spin-charge response: TDDFPT+$U$ treats the linear response of both charge and magnetization densities on equal footing, naturally including their coupling. This avoids the adiabatic decoupling assumption inherent in MFT and $\Delta E$, resulting in a more consistent description of magnetic excitations.
Given these advantages, one might expect TDDFPT+$U$ to yield the most accurate magnon dispersions. Yet, we find that LSWT with $\Delta E$-derived exchange parameters provides slightly better agreement with experimental data. This observation highlights the need for further benchmarking these methods across a broader range of materials to better understand their relative strengths and limitations, but also serves as a reminder of the limitations of the employed energy functionals (hence, the slightly better agreement might be accidental). In particular, using more advanced functionals, such as the extended Hubbard functionals (DFT+$U$+$V$)~\cite{Campo:2010}, which are especially relevant for systems with significant covalent interactions~\cite{Mahajan:2021, Mahajan:2022, Timrov:2022c, Timrov:2023, Gebreyesus:2023, Binci:2023b, Bonfa:2024}, or the orbital-resolved DFT+$U$ formulation~\cite{Macke:2024, Warda:2025}, could change this trend. Specifically, TDDFPT+$U$+$V$ or orbital-resolved TDDFPT+$U$ might ultimately outperform LSWT with $\Delta E$-derived parameters based on DFT+$U$+$V$ or orbital-resolved DFT+$U$, respectively.

We now turn to a more detailed discussion of the $\Delta E$ and MFT methods. Starting with the $\Delta E$ approach: since it computes exchange parameters from total-energy differences between magnetic configurations, accurately resolving very small differences, such as the subtle splitting between nearest-neighbor couplings caused by weak rhombohedral distortion (e.g., $|\mathcal{J}_1^+ - \mathcal{J}_1^-| \sim 0.01$ meV in NiO), requires stringent numerical convergence of DFT+$U$ calculations for large supercells. Although the total energies involved are several orders of magnitude larger than these differences, modern numerical algorithms can achieve the necessary precision. This is analogous, e.g., to the reliable calculation of phonon frequencies using frozen-phonon methods based on supercells and finite differences~\cite{Baroni:2001}. However, achieving high precision convergence in large supercells can be challenging for some systems and computationally demanding, especially as system sizes grow. Regarding this point, our results for NiO and MnO clearly indicate that including exchange interactions up to the fifth nearest neighbor is essential for converged and stable magnon dispersions. However, performing $\Delta E$ calculations for such long-range couplings requires very large supercells containing several hundred atoms, and the associated computational cost scales cubically with system size. This challenge becomes even more significant in materials where long-range and anisotropically split interactions play a key role, such as altermagnets~\cite{Smejkal:2022, Smejkal:2022b}. In these systems, the anisotropic splitting of distant exchange shells must be resolved, for instance, the 7th-nearest-neighbor exchange splitting in MnF$_2$~\cite{Morano:2025} or even the 10th-nearest-neighbor interactions in MnTe~\cite{Liu:2024}. Converging such distant exchange interactions requires very large supercells and high numerical cost, which can make the $\Delta E$ approach increasingly difficult to apply. Despite these challenges, we find that the $\Delta E$ method performs remarkably well in NiO and MnO, accurately reproducing both exchange parameters and magnon dispersions, provided that the Heisenberg model includes a sufficient number of neighboring shells to ensure convergence.

In contrast, the MFT method produces the least accurate results in our study. 
One of the key differences between the $\Delta E$ and TDDFPT+$U$ methods on one hand, and MFT on the other, lies in their treatment of the screening of charge and magnetization densities. Both the $\Delta E$ and TDDFPT+$U$ approaches explicitly incorporate electronic screening effects, whereas MFT neglects them. Specifically, the $\Delta E$ method, as a finite-difference scheme, accounts for screening via self-consistent electronic ground-state minimizations of different spin configurations~\cite{Whangbo:2003, Xiang:2011, Xiang:2013a, Sabani:2020}. The TDDFPT+$U$ method evaluates the dynamical response of charge and magnetization densities fully self-consistently to linear order in the external magnetic perturbation \cite{Gorni:2018, Binci:2025}. In contrast, MFT is a non-self-consistent approach that approximates the screened perturbing potential by the bare external potential, thereby ignoring electronic screening effects~\cite{Korotin:2015, He:2021}. In this context, we note the recent work of Ref.~\cite{Daglum:2025}, which compared self-consistent and non-self-consistent formulations of the spin-spiral method. It was shown that neglecting self-consistency leads to substantial deviations in MIPs and magnon energies relative to TDDFT benchmark data for elemental $3d$ metallic ferromagnets (Fe, Co, and Ni) and selected Heusler compounds. These findings are in line with our observations that the MFT results diverge markedly from those obtained using the self-consistent $\Delta E$ and TDDFPT+$U$ approaches. Therefore, if MFT were extended to include variations of the Kohn-Sham potential induced by infinitesimal spin rotations, its accuracy might approach that of the $\Delta E$ and TDDFPT+$U$ methods. However, it is important to note that the observed deviations of MFT are material-dependent, particularly on the magnitude of the magnetic moments of TM ions and the degree of electron localization versus itinerancy~\cite{Daglum:2025}. Hence, a more systematic investigation across a larger and more diverse set of materials is needed to fully assess the accuracy of the MFT method.

In addition, the lowest accuracy of MFT could be attributed in part to the specific implementation employed, via the TB2J code~\cite{He:2021}, which introduces some approximations. As previously discussed, MIPs obtained via MFT are highly sensitive to technical details of the implementation~\cite{Korotin:2015, MartinezCarracedo:2023, Solovyev:2021, Solovyev:2024}. In particular, Ref.~\citenum{MartinezCarracedo:2023} points out concerns stemming from specific algorithmic choices (we refer the reader to Sec.VII of Ref.~\citenum{MartinezCarracedo:2023} for a detailed discussion). A full analysis of these issues lies beyond the scope of our work, as it would require a deeper examination of the theoretical foundations of MFT. Moreover, recent studies~\cite{Solovyev:2021, Solovyev:2024} highlight that the choice of variables used to represent spin rotations can substantially affect the resulting MIPs, which is another source of variability in MFT-based calculations. Furthermore, as noted earlier, the accuracy of MFT implemented in TB2J for plane-wave-based codes is extremely sensitive to the quality of the wannierization, particularly the atom-centering of the resulting Wannier functions. Hence, to further assess the accuracy of MFT, alternative wannierization strategies could be explored, such as symmetry-adapted Wannier functions~\cite{Pizzi:2020} or even approaches that avoid Wannier functions altogether, as in the recently proposed method of Ref.~\cite{Skovhus:2025}. Finally, using alternative MFT implementations, for instance based on KKR~\cite{Luders:2001, Papanikolaou:2002} or LMTO~\cite{Andersen:1984, Turek:1997}, could provide further insights into the accuracy and reliability of the MFT method. However, a major challenge is that these approaches rely on different DFT codes and use distinct levels of theory for the ground-state description, which differ from ours and complicate the direct comparison of computed MIPs and magnon dispersions~\cite{Fischer:2009, Solovyev:1998}.

\section{Conclusions}
\label{sec:conclusions}

Accurate modeling of magnetic excitations in transition-metal compounds remains a challenge for DFT-based approaches. In this work, we employed a consistent DFT+$U$ ground state for all calculations, with the Hubbard $U$ parameter computed from first principles via DFPT~\cite{Timrov:2018}. This eliminates ambiguities related to empirical $U$ choices and allows a focused evaluation of each method's accuracy in computing MIPs and magnon dispersions based on their underlying physical approximations.

First, we showed that direct methods for extracting exchange parameters require careful convergence tests with respect to the interaction range in the Heisenberg Hamiltonian~\eqref{eq:heisenberg_hamiltonian_0}. For both NiO and MnO, we found that including exchange interactions up to the fifth nearest neighbor is needed when using the MFT or $\Delta E$ methods; otherwise, unphysical features such as magnon instabilities may arise (see e.g. the $\Delta E$-derived magnon dispersion of NiO in Fig.~S1 in the SM~\cite{SupplementalMaterial}).
Additionally, we demonstrated the importance of including the rhombohedral lattice distortion, which splits the nearest-neighbor interaction into two distinct parameters, $J_1^+$ and $J_1^-$, due to magnetoelastic coupling. This splitting is crucial for reproducing the finite magnon energy at the M point in the BZ.

Next, we computed magnon dispersions using LSWT based on first-principles $J$ parameters. Among all methods, LSWT based on MFT-derived parameters showed the poorest agreement with experimental inelastic neutron scattering data for both NiO and MnO. In contrast, LSWT with $\Delta E$-derived exchange parameters and TDDFPT+$U$ both yield good agreement with the measured magnon dispersions, with the former providing the best overall accuracy. To further analyze these differences, we employed a minimal Heisenberg model containing only three effective exchange parameters: $\mathcal{J}_1^+$, $\mathcal{J}_1^-$, and $\mathcal{J}_2$, obtained by fitting to the previously computed magnon dispersions as well as to experimental data. We found that $\Delta E$-fitted effective parameters match the experimental values most closely, while MFT-fitted values show the largest deviations. TDDFPT+$U$-fitted parameters are slightly less accurate than $\Delta E$-fitted ones, but crucially, they reproduce the correct trends among $\mathcal{J}_1^+$, $\mathcal{J}_1^-$, and $\mathcal{J}_2$, consistent with experiment for both materials.

Our comparative study provides a valuable contribution to the ongoing efforts aimed at assessing the accuracy of first-principles methods for describing magnetic interactions in transition-metal compounds. While our analysis focused on two prototypical systems, NiO and MnO, a broader benchmarking across a wider range of materials, with varying crystal structures and chemistries, is essential to draw more general conclusions. Each method considered in this work, $\Delta E$, MFT, and TDDFPT+$U$, has its own strengths and limitations in terms of accuracy, robustness, and computational cost. The $\Delta E$ and MFT approaches are less computationally expensive than TDDFPT+$U$, making them more suitable for large or complex systems; however, their accuracy depends on the number of exchange parameters included (see Fig.~S1 in the SM~\cite{SupplementalMaterial}) and, in the case of MFT, on the implementation details. In contrast, TDDFPT+$U$, while significantly more expensive in terms of computational resources, offers an accurate and physically consistent framework that does not rely on an arbitrary selection of exchange parameters in the underlying model. Moreover, the computational complexity of the three methods, briefly summarized in Sec.~S3 of the SM~\cite{SupplementalMaterial}, differs significantly: The $\Delta E$ method relies on a series of standard self-consistent DFT+$U$ supercell calculations in which selected magnetic moments are flipped relative to the ground state; The MFT approach is a perturbative method based on the Green’s function formalism and, in certain implementations, additionally requires wannierization; TDDFPT+$U$ involves the explicit solution of linear-response equations, which substantially increases both the computational time and memory requirements. Moreover, extending TDDFPT to more advanced functionals (e.g., DFT+$U$+$V$ or hybrid functionals) is theoretically and technically more involved. Therefore, while a single method can provide useful insights, predictive modeling of magnetic excitations benefits significantly from the use of multiple approaches. Whenever feasible, using multiple computational approaches provides a more reliable and comprehensive understanding, particularly when interpreting or guiding experimental investigations. Looking ahead, the continuous development and systematic benchmarking of these methods across diverse material classes will be crucial to advancing our ability to predict and understand magnetic excitations from first principles with high confidence.

\section*{ACKNOWLEDGEMENTS}

F. dos S. thanks Manuel dos Santos Dias for fruitful discussions. F. dos S., L.B., N.M., and I.T. acknowledge support from the NCCR MARVEL, a National Centre of Competence in Research, funded by the Swiss National Science Foundation (Grant No.~205602). I.T. acknowledges support from the Swiss National Science Foundation (Grant No.~200021-227641 and No.~200021-236507). L.B. acknowledges the Fellowship from the EPFL QSE Center ``Many-body neural simulations of quantum materials'' (Grant No.~10060). G.M. acknowledges support by the MUR through the PRIN  project ``q-LIMA'' (Grant No.~2020JLZ52N). This work was supported by a grant from the Swiss National Supercomputing Centre (CSCS) under project ``lp18'', ``s1326'' and ``mr33''. The computational resources have been provided by \emph{computing@unipi}, a computing service provided by the University of Pisa, the HPC center of the Fondazione Istituto Italiano di Tecnologia (IIT, Genova). G.M. also acknowledges the CINECA award under the ISCRA initiative, for the availability of high-performance computing resources and support, in particular through the ``ISCRA C'' projects ``HP10CAVD6L'', ``HP10CLNRW9'', and ``HP10CY46PW''.


%

\end{document}